\documentclass[aps,prl
,preprint
,endfloats*
,groupedaddress,showpacs]{revtex4-1}
\usepackage[]{graphicx}
\begin{document}
\title{Structural behavior of Pb$_{y}$Bi$_{1.95-y}$Sr$_{1.49}$La$_{0.4}$Cu$_{1.15}$O$_{6+\delta}$ for 0$<$y$<$0.53}
\author{Olaf L\"ubben}
\email[]{luebben@physik.hu-berlin.de}
\author{Lenart Dudy}
\author{Alica Krapf}
\author{Christoph Janowitz}
\author{Recardo Manzke}
\affiliation{Humboldt Universit\"at, Institut f\"ur Physik, Newtonstrasse 15,
12489 Berlin, Germany}


\date{\today}

\begin{abstract}
In the Bi cuprates, the presence of a near 1$\times$5 superstructure is well
known. Usually, this superstructure is suppressed by the
substitution of lead, but there have been reports of a phase separation in so
called $\alpha$ and $\beta$ phases. This paper shows in high detail time how and
 why the phase separation develops and what happens to the quasi-1$\times$5 
superstructure upon lead substitution. For this purpose, the lanthanum- and lead-substituted single-layered superconductor Bi$_{2+z}$Sr$_{2-z}$CuO$_{6+\delta}$ 
has been investigated by 
scanning tunneling microscopy and low-energy electron
diffraction. The La content was kept constant at slightly under-doped
concentration while the Pb content was changed systematically. Thermodynamic
considerations show that a phase mixture of $\alpha$ and $\beta$ phases is
inevitable.
\end{abstract}


\maketitle
\section{Introduction}
For surface-sensitive experiments probing the high-temperature superconductivity
of the hole-doped cuprates, single crystals of the bismuth cuprate family are
the preferred choices because of the natural cleaving plane between two abreast
van der Waals binded BiO layers. Preferentially used in experiments is the
compound Bi$_{2+z}$Sr$_{2-z}$CaCu$_2$O$_{8+\delta}$ (BISCO or Bi2212) with two
CuO$_2$ layers per unit cell. Since a strong interlayer interaction of next
nearest CuO$_2$ planes leads to split bands in this Bi2212 (see, e.g.,
Refs. \cite{Kordyuk2002} and \cite{Asensio2003}), other members of the family have come into the
focus of research. Bi$_{2+z}$Sr$_{2-z}$CuO$_{6+\delta}$ (Bi2201) with only one
CuO$_2$ layer per unit cell is often regarded as the most ideal experimental
manifestation of the single undisturbed CuO$_2$ plane. By controlling the extra
oxygen ($\delta$) and varying the amount of Bi in Sr positions ($z$), the
hole doping can be tuned around optimum doping. Furthermore unlike for Bi2212,
nearly the whole range of hole dopings can be achieved due to substituting Sr by
La: Bi$_{2+z}$Sr$_{2-z-x}$ La$_x$CuO$_{6+\delta}$ (La-Bi2201).

Structurally the Bi cuprates are far from ideal crystals. Since they consist of
alternating rocksalt and perovskite building blocks they are prone to
superstructures which release the stress. For instance Bi2212 and La-Bi2201
exhibit a pronounced structural imperfection -- an incommensurate
superstructure in $b$ and $c$ directions. This superstructure is sometimes called
``Bi-type superstructure'' \cite{Zhiqiang1993}, or is referred by its in-plane
periodicity as the ``near 1$\times$5 superstructure''. For La-Bi2201, it can be
shown that this superstructure can be monoclinic or orthorhombic
\cite{Martovitsky2007} -- depending on the La content. However, the question
of the driving force for the superstructure in the Bi cuprates gave rise to
basically two models, the ``extra oxygen model'' and the
``misfit-model'' (see, e.g., Refs. \cite{Zhiqiang1993} and \cite{Zhiqiang1997, Zandbergen1988,
Zandbergen1990}). Besides the strive for a structurally pure crystal, the
suppression of this near 1$\times$5 superstructure is highly desirable for
other reasons. It has a periodicity around 27\,\AA{}, comparable to the in-plane
superconducting coherence length of approximately 30\,\AA{} (see, e.g., Refs. 
\cite{Palstra1988} and \cite{Vedeneev1999}); and for experiments probing the reciprocal space, the superstructure hides intrinsic features. This attribute is
exemplified in angular-resolved photoemission spectroscopy. Diffraction replica of the outgoing electrons are produced,\cite{Aebi1994,Ding1996} which mask the true electronic structure.

The suppression of the quasi-1$\times$5 superstructure by substituting a certain
fraction of Bi by Pb is a method often used for the Bi cuprates, leading to
Pb$_y$Bi$_{2+z-y}$Sr$_{2-z-x}$La$_x$CuO$_{6+\delta}$ [(Pb,La)-Bi2201],
Pb$_y$Bi$_{2+z-y}$Sr$_{2-z}$CuO$_{6+\delta}$  (Pb-Bi2201), and
Pb$_y$Bi$_{2+z-y}$Sr$_{2-z}$CaCu$_2$O$_{8+\delta}$ (Pb-Bi2212). Besides the
suppression of the superstructure, the substitution by Pb is also known to
enhance
$T_C$, i.e.\,, from about 29\,K for optimally doped La-Bi2201 to about 38\,K for
optimally doped (Pb,La)-Bi2201. However, examinations of Bi2201 (Ref.\,\cite{Zhiqiang1993}) and also La-Bi2201,\cite{Zhiqiang1997} showed that at
intermediate Pb concentrations, a ``Pb-type'' superstructure replaces the
quasi-1$\times$5 superstructure. For the two-layer material Pb-Bi2212, it was
reported \cite{Chong1997} that at a comparable intermediate Pb level, a phase
separation in so-called $\alpha$ and $\beta$ phases occurs. The $\alpha$ phase
was found to have high corrugation and to be Pb poor, whereas the $\beta$
phase was found to have low corrugation and to be Pb rich.\cite{Hiroi1998}
Therefore, this topology was examined in
Bi$_{2-y}$Pb$_y$Sr$_2$CaCu$_2$O$_{8+\delta}$ by using a focused electron beam
(30\,nm in diameter) which was produced by a high-resolution electron microscope. For the one-layer material Pb-Bi2201, a quite similar phase
separation was recently shown.\cite{Nishizaki2007}
The present study deals with well-characterized single crystals of a systematic
series of the lead- and lanthanum-substituted single-layer (Pb,La)-Bi2201, where
the Pb concentration was varied continuously between 0 and 0.53 f.u.
The lanthanum content was kept constant. These crystals were probed by low
energy-electron diffraction (LEED) and topographical scanning tunneling
microscopy (STM). Hiroi et al.\,\cite{Hiroi1998} proposed a phase diagram for
Pb-Bi2212 to describe the appearance of $\alpha$ and $\beta$ phases upon
varying the Pb content. It will be shown, that the one-layer (Pb,La)-Bi2201
probed here also shows intermediate phases and has a similar phase
diagram. In this paper, the disappearance of the near 1$\times$5
superstructure, the appearance of intermediate $\alpha$* and $\beta$* phases,
and of the final $\alpha$ and $\beta$ phases are discussed in detail for
(Pb,La=0.4)-Bi2201. Based on the results a general advanced phase
diagram will be proposed. 

\begin{figure}
\includegraphics[width=\columnwidth]{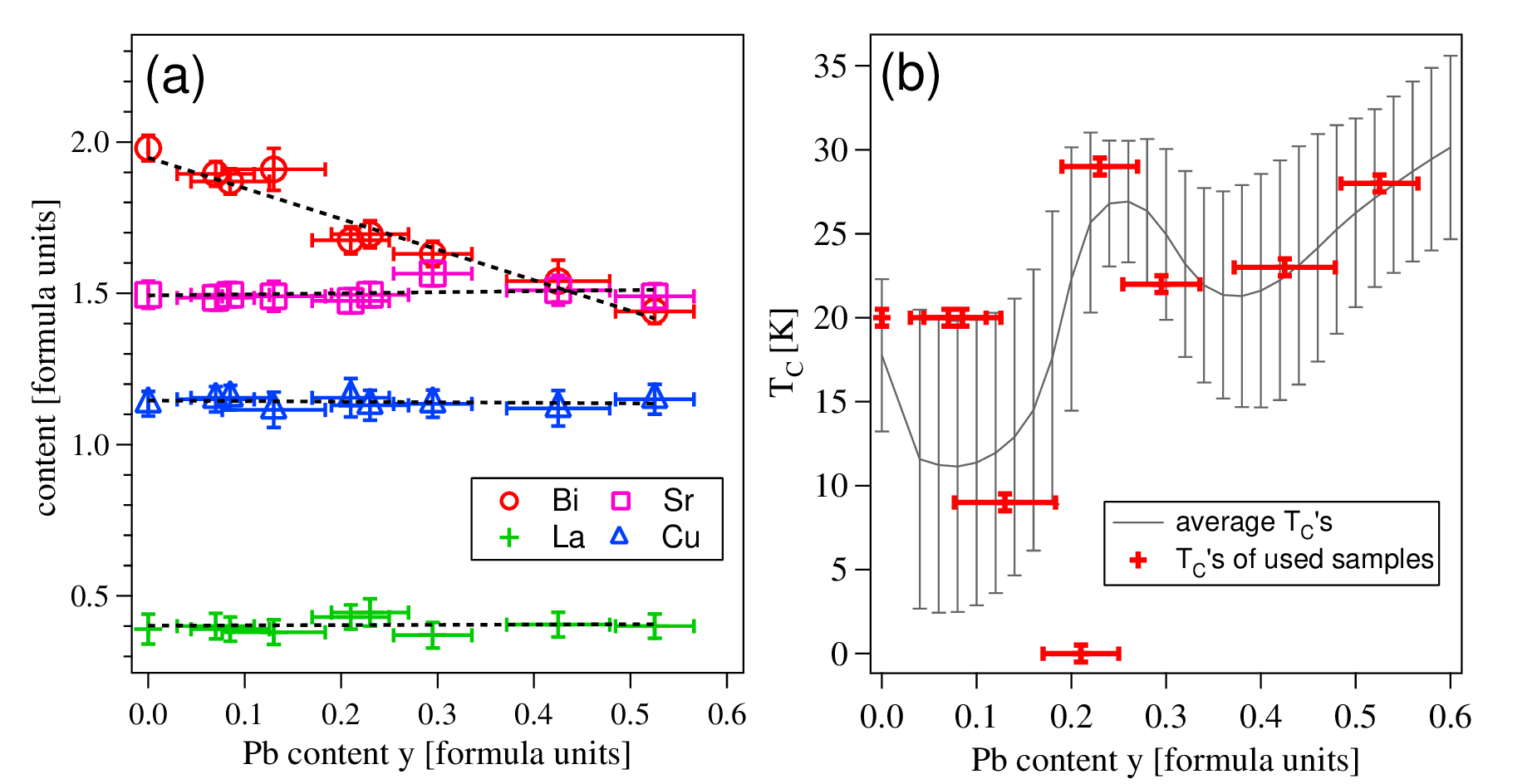}%
\caption{\label{FIG1}(Color online) (a) Stoichiometric composition of the
samples. The Bi,
Sr, Cu, and La content relative to the Pb substitution level is shown together
with a line fit (dotted lines) for each element as a function of the Pb content.
The resulting formula confirms the vanishing influence of the Pb on the Cu and
Sr content: Pb$_y$Bi$_{1.95-y}$Sr$_{1.49}$La$_{0.4}$Cu$_{1.15}$O$_{6+\delta}$.
(b) Superconducting transition temperature $T_C$ relative to the Pb substitution
level of the samples used in this study (crosses with error bars). The thin
curve was calculated by averaging over many samples and represents the average
$T_C$ and its standard error, indicated by bars.} 
\end{figure}

\section{Experimental}
Single crystals of (Pb,La)-Bi2201 were grown using the flux method. For the
flux, Bi is used in typical additional amounts of 10\,\% of the total Bi formula
unit. The dependence between nominal and actual La composition is similar as in
Yang et al.\cite{Yang1998} The starting mixture is composed of the oxides
of Bi$_2$O$_3$, CuO, LaO, PbO, and the carbonate SrCO$_3$, which are 99.98\,\%
pure. To get a homogeneous mixture, the constituents were dissolved in ethanol
and then ground. After this, it was calcined at $T\approx$800$^{\circ}$C. For
crystals containing lead, first the oxides Bi$_2$O$_3$, CuO, and the carbonate SrCO$_3$ are
mixed. After the calcination, LaO and PbO are added. Then, the composition is
heated in a zirconium-oxide crucible above the liquidus temperature, which is,
dependent on the exact composition, between 950 and 1050$^{\circ}$C. This
temperature is maintained for 1 to 2\,h to thermally homogenize the melt. After
the homogenization, the system is cooled. In the temperature range roughly
between 800 and 880\,$^{\circ}$C, the system is cooled gradually at a rate of
1--2\,K/h to allow the formation of only a few crystallization nuclei. It is then
rapidly cooled to room temperature (RT) with a typical rate between 10
and 20\,K/h. After preparing single crystals out of the crucible, the crystals
were characterized by ac-susceptibility measurements, to obtain the transition
temperature $T_C$, and energy dispersive x-ray analysis to determine the
actual chemical composition. Figure \ref{FIG1}(a) shows the obtained chemical
composition dependent on the Pb content. By the line fits (dotted lines), the
series can be described as
Pb$_y$Bi$_{1.95-y}$Sr$_{1.49}$La$_{0.4}$Cu$_{1.15}$O${6+\delta}$.
Figure \ref{FIG1}(b) shows the $T_C$ dependent on the Pb content. The thick crosses indicate the individual samples, whereas the black line represents the
average of many samples. Only for the sample with $y=0.21$, does the $T_C$ differ
significantly from the area spanned by the standard error of the averaged curve.
Topological scanning tunneling microscopy was done using a systematic series of
these crystals. All measurements were conducted at room temperature in
ultra-high vacuum at a pressure of $<1\times10^{-10}$\,mbar immediately after cleavage in
situ. The analysis of the surface structure was performed by a commercial
Omicron STM (VT 25 SPM). The tips were either made from electrochemically etched
tungsten wire or pinched-off PtIr wire.

\begin{figure*}
\includegraphics[width=\textwidth]{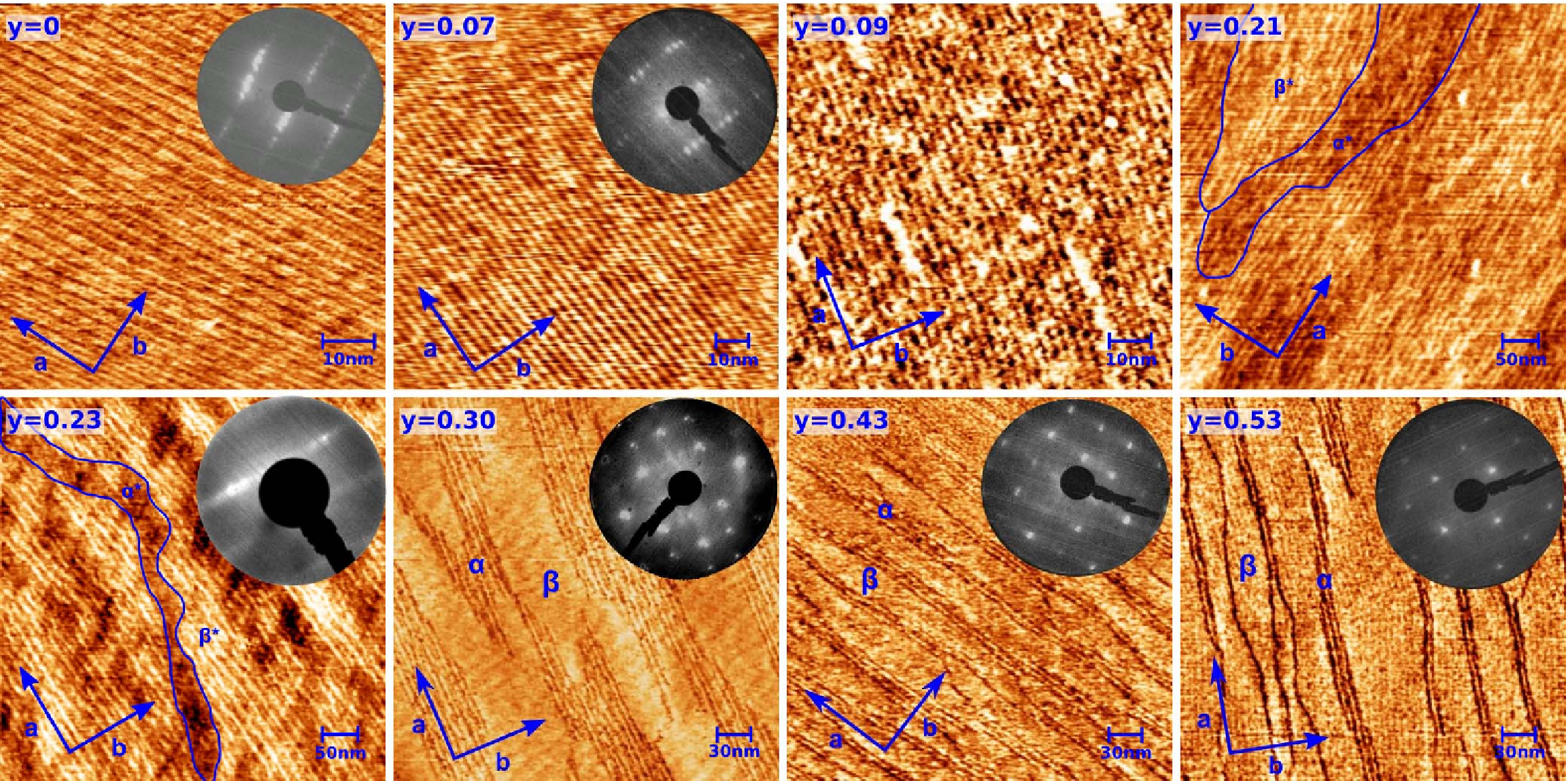}%
\caption{\label{FIG2}(Color online) Representative, (equally oriented) STM and
LEED patterns
of the samples with Pb substitution of 0, 0.07, 0.09, 0.21, 0.23, 0.30, 0.40,
and 0.53 f.u.. The Pb substitution level is indicated in the upper left
of each pattern. In the lead free crystal in the upper left ($y=0$), the
1$\times$5 superstructure along the $b$ direction can be seen as a striped
modulation in the STM pictures and a line of additional reflexes in the LEED
pattern. At $y=0.07$, the Pb substitution causes extra bright spots in the
STM picture. At $y=0.09$, these bright spots begin to form bright linear objects.
At intermediate Pb levels ($y=0.21$, 0.23), the existence of bright linear objects
is clearly visible in the STM pictures. These bright linear objects seem to self
organize both in regions with higher density ($\beta^*$ phase) and with lower
density ($\alpha^*$ phase). In the LEED picture ($y=0.23$), a blurred structure
indicates still modulations in the $b$ direction. At higher Pb levels ($y=0.30$,
0.40, and 0.50), the STM pictures show a clear phase separation in an $\alpha$ and an
$\beta$ phase. The fraction of the $\alpha$ phase reduces with increasing
Pb level. The quasi-1$\times$5 superstructure or its remnants are absent in the
LEED
pictures for Pb concentrations of 0.30 and higher. When going from $y=0.30$ to
$y=0.53$, the LEED spots get sharper and the incoherently scattered background
between the spots is reduced.} 
\end{figure*}	

In the STM image for the Pb-free samples (Fig.\,\ref{FIG2}, $y=0$), the quasi-1$\times$5 superstructure is visible as a striped modulation. This can also be
seen clearly by the extra spots in the LEED pattern. The substitution with
y=0.07 f.u.\,leads to some brighter dots in the STM image. These
brighter dots represent a higher unoccupied density of states and are oriented
along the
superstructure. With reference to other STM measurements for Pb-Bi2201
(Ref.\,\cite{Mashima2006}) and for Pb-Bi2212,\cite{Kinoda2003} it is assumed that these
brighter dots are the substituted Pb atoms. The superstructure is present in the
STM image and as well in the LEED picture. When slightly increasing the Pb
substitution ($y=0.09$) the brighter dots, which were already visible in the
$y=0.07$ sample, cluster along the present superstructure. These linear objects
maximize their mutual separation which seemingly corresponds to the
most energetically favorable position. In the STM image for a sample with Pb
substitution of $y=0.21$, the density of these linear objects is significantly
increased compared to the sample with $y=0.09$. It is remarkable that these
objects themselves cluster, denoted as the $\beta$* phase. The clustered objects
are separated by a second phase, denoted here by $\alpha$*. This phase has a
lower density of bright linear objects with washboard like appearance. Since the
bright objects may represent ordered Pb atoms, the $\alpha$* phase is a low Pb
phase, compared to the $\beta$* phase. At substitution of $y=0.23$, the linear
objects within the $\beta$* phase are more pronounced. The near 1$\times$5
superstructure is still detectable by the blurred lines in the LEED pattern but
weak. The sample with Pb content of $y=0.3$ shows smooth and bright areas without
any linear structure. It is assumed that this $\beta$ phase is Pb rich, which is
supported by a report of Hiroi et al.,\cite{Hiroi1998} where this topology
was examined in Pb-Bi2212 by using a focused electron beam (30\,nm in diameter)
which was produced by a high-resolution electron microscope. Similar as in the
report of Hiroi et al., beneath the flat $\beta$ phase, a second, washboard-like
$\alpha$ phase shows up. This $\alpha$ phase consists of linearly ordered,
equidistantly spaced bright objects. The extra spots in the LEED picture are
gone, indicating that a quasi-1$\times$5 superstructure is no longer
detectable. Increasing the Pb substitution further to, e.g., $y=0.43$ or $y=0.53$
results in no new topological features. The STM image consists of the
$\alpha$ and the $\beta$ phase. The fraction of the $\alpha$ phase decreases
enhancing the fraction of the $\beta$ phase, when the proportion of Pb is
increased.
The near 1$\times$5 superstructure remains absent in the LEED picture.
Furthermore, the LEED spots get sharper and the incoherently scattered
background between the spots is reduced.

\begin{figure}
\includegraphics[width=\columnwidth]{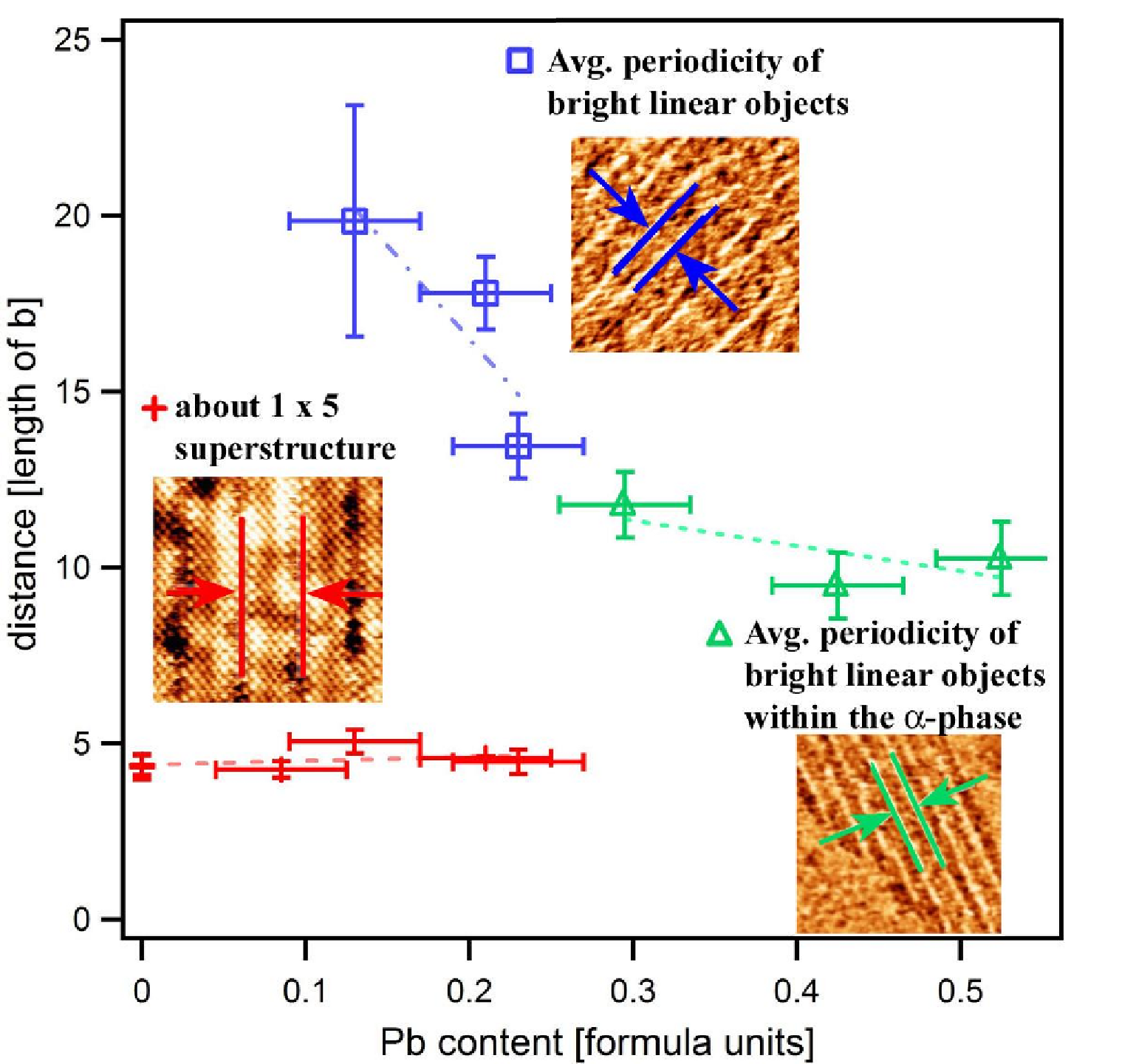}%
\caption{\label{FIG3}(Color online) Quantitative lengths of the
superstructures and features
from self-organization obtained by the average from many STM pictures. The 
characteristic length produced by the near 1$\times$5 superstructure is shown
by red crosses. This superstructure wavelength remains nearly constant for all
Pb substitution levels. A second characteristic length is shown quantitatively
as blue squares. It is the average length between the bright linear objects at
$y=0.09$, $y=0.21$, and $y=0.23$ Pb concentration (to be compared with Fig.\,2). For
higher Pb substitutions ($y\geq0.3$), a third characteristic length can be
established as the average length between the bright 
linear objects. It is depicted as green triangles which are taken solely from
the $\alpha$ phase since the smooth $\beta$ phase shows no modulations any
more.} 
\end{figure}

In Fig.\,\ref{FIG3}, the quantitative analysis of the found distances by the
(near 1$\times$5) superstructure and the bright objects are shown. To do this
the periodicity was averaged over many patterns. For computing the length in
units relative to $b$, the absolute length was divided by 5.4\,\AA{}. From this
graph it is clear that the near 1$\times$5 superstructure remains nearly
constant and vanishes upon the appearance of the $\alpha$- and $\beta$-phase
separation. Please note that this is in contradiction to other results on
Pb-Bi2201,\cite{Nishizaki2007} where the superstructure wavelength increases
upon increasing the Pb level. Here, the distance between the bright objects
reduces when increasing the Pb substitution level and its length scale seems to
evolve into a constant value of quasi-1$\times$10.

\section{Discussion}
\begin{figure}
\includegraphics[width=\columnwidth]{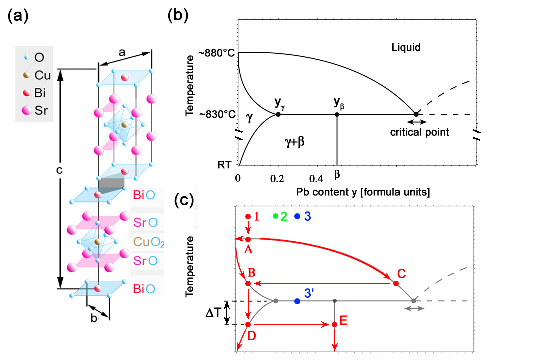}%
\caption{\label{FIG4}(Color online) (a) Ideal structure of Bi2201. The
structure can be seen as
built from a perovskite block, containing the CuO$_6$ octaeder and the Sr atoms,
expanded in $c$ direction by a rocksalt component which is by the BiO and
SrO layer. This arrangement is repeated in $c$ direction with a shift of $a/2$. As
the substitution of Bi by Pb leaves the perovskite component nearly unchanged
(see text), the substitution can be seen as facilitating the adaption of the
BiO plane relative to the perovskite block. (b) Pseudobinary phase diagram,
dependent on temperature and Pb level: At low Pb substitutions and temperatures,
there is the trianglelike-shaped $\gamma$ phase, which is the phase with the
1$\times$5 superstructure. Beginning at low temperatures toward higher levels
of Pb substitutions,
there is an area denoted as $\gamma$+$\beta$, where a mixture between the
$\gamma$ phase and the $\beta$ phase occurs. The $\beta$ phase itself is assumed
to occur at the specific Pb substitution level where the (Bi,Pb)O plane is
perfectly adapted on the perovskite block. This Pb level is assumed to be 0{.}5.
This optimally adapted $\beta$ phase is shown as a vertical line marked with
$\beta$ at the abscissa. The stability of the $\beta$ phase ends at the point
y$_{\beta}$, at the horizontal eutectic line. This line is situated at a
temperature of about $T=830^\circ C$ and starts from the point y$_{\gamma}$,
where the $\gamma$ phase forms, and goes at least to the eutectic or
peritectic point to the right, which is the unmarked crossing point marked with
a horizontal arrow below. Above this point is the liquid (melt). Left from this
point at temperatures higher than 830$^{\circ}$C is another region of phase mixture,
where the liquid phase coexists with a solid phase. (For simplification, the
notation $\gamma$+L for this phase is omitted here.) (c) Crystallization process
upon cooling within the phase diagram to specify and explain the certain
observed morphologies. There are three points, marked as 1, 2, and 3, which represent
preeminent concentrations. As an example, the red arrows and circles depict the
cooling/crystallization process for composition 1. For details on this and the
other points 2 and 3 see the text, where also the blue point, marked with 3', is
discussed.} 
\end{figure}

From the data, it is evident that a multitude of morphologies exist. Therefore,
the most relevant mechanisms responsible for the morphologies, dependent on Pb
substitution, need to be addressed. The ideal structure of Bi2201, as depicted
in Fig.\,\ref{FIG4}(a), can be seen as built from a perovskite block, containing
the CuO$_6$ octaeder and the Sr atoms, expanded in $c$ direction by a rocksalt
component, built by the SrO and BiO layer. 
Three parameters affect the structural distortion in this
system, these are the extra oxygen, the substituted lanthanum, and the misfit of
the BiO plane to the perovskite block.

The extra oxygen and the lanthanum are dopant atoms, meaning that they change
the binding conditions in the CuO$_2$ plane. This is due to the Pauling-type
contraction of the CuO binding, which decreases the planar binding length with
increasing hole doping.\cite{Roehler2005} It is worth noting that although
the Pb substitution is changed, the hole doping is constant in the probed
series. The hole concentration is given as $p=0.24-0.21x$ at Pb content $y=0$
 (Ref.\,\cite{Schneider2005}) while for Pb content $y=0.4$ it is $p=0.23-0.15x$,\cite{Ariffin2009} where $x$ denotes the lanthanum content. This means that
changing the Pb content from 0 to 0.4 f.u.\,at constant lanthanum of
$x=0.4$ would result only in a marginal hole concentration difference of about
0.015 holes per Cu atom. By keeping the hole concentration approximately on the
same level, the CuO bond length was kept constant. The perovskite block can
therefore be regarded as static in the whole probed series and cannot be the
reason for the change in morphology. Thus, the driving force in the change in
morphology is suggested to be the mismatch of the (Bi,Pb)O layer and the static
perovskite block.

Following this line of argument, the $\beta$ phase corresponds to the case,
where the (Bi,Pb)O plane is perfectly adapted on the perovskite block. The STM
patterns suggest that this adoption will occur at about 0.5 f.u.\,of Pb
substitution. Upon increasing Pb substitution, the (Bi,Pb)O layer has its average
size determined by the Pb to Bi ratio. Therefore, to account for the
morphologies here, we can design a pseudo-binary phase diagram dependent on the
Pb substitution $y$. Similar morphologies in the two-layer material suggest that
the
diagram only needs to be slightly modified to account for Pb-Bi2212. The phase
diagram is depicted in Fig.\,\ref{FIG4}(b). The vertical axis marks the
temperature and starts at room temperature. The Pb substitution level is on
the horizontal axis.

At low Pb substitution, the $\gamma$ denoted phase is present. This phase is the
lead-free crystal with lanthanum substitution of 0.4, which exhibits the quasi-1$\times$5 superstructure. It crystallizes at about 880$^{\circ}$C. The optimally
adapted smooth $\beta$ phase is realized at around $y\approx0.5$ Pb content and
crystallizes approximately at 830$^{\circ}$C. 
The shown phase diagram is further characterized by a eutectic or peritectic
point. As the arrow implies, the exact position of this point is not certain.
Further studies might be necessary to find its exact position. This
uncertainty is of minor importance with regard to the
morphology as seen in STM up to a Pb concentration of approximately
$y\approx0.5$. In principle, the optimally adapted $\beta$ phase could be
peritectic as drawn or eutectic. 
Peritectic means here that the
$\beta$ phase cannot be directly reached by cooling the melt. In the eutectic
case, the critical point would be at the same
Pb concentration as $\gamma_{\beta}$, which would mean
that the phase can be reached directly by cooling from the melt.
It should be noted that there is no $\alpha$ phase in the
diagram. 

We begin with a crystallization process from the melt at point 1 in
Fig.\,\ref{FIG4}(c), corresponding to a low Pb concentration. Lowering the
temperature at this composition splits the system at the liquidus into two
phases: one part crystallizes in the $\gamma$ phase [at point A in
Fig.\,\ref{FIG4}(c)] whereas the rest of the system follows the liquidus
toward the eutectic point. A gradual reduction in the temperature will enable
the $\gamma$ phase to grow at the cost of the melt. This growth can continue
until point C is reached and the concentration in $\gamma$ is identical to the
starting composition (point B). Lowering the temperature further leaves the
system stable within the $\gamma$ phase until point D is reached. At this point,
the $\beta$ phase decomposes from the $\gamma$ phase. Because the system is
strongly supercooled [$\Delta T$ in Fig.\,\ref{FIG4}(c)], the decomposition of
the $\beta$ phase will be sudden. The Pb atoms crystallize at local
inhomogeneities (see, $y=0.07$ and 0.09 in Fig.\,\ref{FIG2}). These local
inhomogeneities are, for example, given by the superstructure. Therefore, in the
STM images bright spots are visible, which are located near the
superstructure. 

Increasing the starting Pb concentration (point 2) will
increase the amount of condensed Pb near the superstructure. 
As composition 2 is reached, the $\beta$ phase can be stabilized without
supercooling. But the stabilization without supercooling will only be possible
under adiabatic conditions. Two topologies will be typically formed (see $y=0.21$
and $y=0.23$ in Fig.\,\ref{FIG2}). One topology is the very locally stabilized
$\beta$ phase ($\beta$* phase) and the other topology is quite similar to the
washboard like structure from supercooling ($\alpha$* phase). The existence of
large quantities of the $\beta$ phase is dependent on the cooling rate and
the Pb level. The lower threshold for the occurrence of large fractions of the
$\beta$ phase in the crystals is at around $y\approx0.3$ for the applied growth
conditions (see $y=0.3$ in Fig.\,\ref{FIG2}). At this composition, the character
of the $\alpha$ phase becomes clear. The $\alpha$ phase is a phase mixture
produced by the part of the system which crystallizes at point $y_{\gamma}$ into
the $\gamma$ phase and decomposes upon cooling further in $\gamma$
and $\beta$ phases. This explanation of the $\alpha$ phase as a phase mixture is
quite interesting, as it means that in a sample at any lead concentration
between $y_{\gamma}=0.2$ and $y_{\beta}=0.5$, two (metastable) phase mixtures
develop.

From point 3' in Fig.\,\ref{FIG4}(c) on the eutectic line,
which is reached from point 3 by cooling, the Pb-rich $\beta$ phase grows 
until point y$_{\beta}$, where upon cooling the pure $\beta$ phase
crystallizes. All material not crystallizing in this way moves from point 3' to
the left along the eutectic line to the $y_{\gamma}$ point and further upon
cooling to RT along the $\gamma-(\gamma+\beta)$ phase boundary. The excess
Pb not used in crystallization along the first way determines the
fraction of $\beta$ phase in the $\gamma$ phase forming the pseudo-phase
$\alpha$. The decomposition between $y_{\beta}$ and $y_{\gamma}$ makes
$\beta$  and $\gamma$ phases align themselves perpendicular to the growth
direction (see also Ref.,\cite{Hiroi1998}), which is for Bi2201 in the $a$
direction. The
washboard-like structure in the $\alpha$-phase mixture should order at local
inhomogeneities, present in the already crystallized $\gamma$ phase. 

From this explanation of the observed morphologies by general
thermodynamic interpretations, clear conclusions can be drawn. First of all, a
multitude of topological morphologies due to lead substitution will occur for
any crystal-growth method, e.g., for the flux method, as in our case, but also
for the traveling solvent floating-zone method (see,
e.g., Ref.\,\cite{Nishizaki2007}).  Furthermore, as already reported by Hiroi et
al.,\cite{Hiroi1998} a quite similar phase diagram leads to similar
morphologies for the double CuO$_2$-layered Pb-Bi2212. In other words, in most
conventional cases, up to the Pb level examined here, the occurrence of a
minority washboard-like $\alpha$ phase cannot be avoided in Pb-substituted
bismuth cuprates.

The large average spacing between the $\alpha$-phase regions leads to short
reciprocal vectors in the $b$ direction. For a Pb concentration of 0.4,
for instance, it is 1$\times$32 (Ref.\,\cite{Dudy2007}) but this of course varies with
the amount of lead and with the cooling rate. As shown in Fig.\,\ref{FIG3}, the
bright linear objects forming the washboard-like structure within the $\alpha$
phase converge to an average distance of about 10$\times$b for high Pb
concentrations of around 0.4. It can be suggested that these distances represent
the so-called additional Pb-type modulation, visible in electron-diffraction
experiments by Zhiqiang et al.\,\cite{Zhiqiang1993,Zhiqiang1997} Its
oscillation strength reduces considerably with increasing Pb content, as the
$\alpha$ phase becomes the minority phase with only few linear
objects within a typical $\alpha$-phase area. 

We should now discuss how the above-mentioned modulations
influence the LEED pictures. While in Fig.\,\ref{FIG2} for $y=0$ and $y=0.07$, the
quasi-1$\times$5 shows up as clear, separated spots in LEED, for $y=0.23$ the
disturbed morphology leads to a diffuse LEED pattern along the $b$ direction.
For $y=0.3$ and $y=0.43$, the fundamental reflexes are of dominating intensity. The
diffuse intensity between these main spots is ascribable to the short-scaled
reciprocal vectors from the $\alpha$-$\beta$ phase separation, which is not
individually resolvable anymore. For $y=0.53$, the spots are sharp and the diffused
part is minimized although the alpha phase is still present. This can be
understood as follows:
the $\alpha$ phase only occupies a small fraction of the surface and the number
of bright
linear objects in it is just between one and three, giving therefore low
oscillation strength from the 1$\times$10 periodicity. Also, the spacing between
the $\alpha$-phase regions is beyond 1$\times$32 yielding reciprocal vectors
and outside the coherence length of the LEED experiment. However, other experiments
probing the reciprocal space, e.g., angle resolved photoemission, could be influenced by these structural details.

In addition, it is reasonable to suggest that the superconducting
properties are affected by highly inhomogeneous samples. This structural
impurity might therefore be the reason for the discontinuity of the
superconducting temperature between 0 and
0.3 f.u.\,of lead (see Fig.\,\ref{FIG1}). In this region, a mixture of
different crystalline phases exists which are highly inhomogeneous (as can be
seen in Fig.\,\ref{FIG2}, $y=0.09$, 0.21, and 0.23). This disturbed transition phase
of
the crystal might have an influence on the formation of Cooper pairs in
the CuO$_2$ layer since any structure shorter than the coherence length is
detrimental.

\section{Conclusion}
We have studied in detail the morphology of
Pb$_y$\-Bi$_{1.95-y}$\-Sr$_{1.49}$\-La$_{0.4}$\-Cu$_{1.15}$\-O$_{6+\delta}$ for
$0\leq y\leq 0.53$ by LEED and STM. Upon increasing the Pb substitution, the
well-known quasi-1$\times$5 superstructure vanishes and other phases occur.
Above $y=0.3$, a flat Pb-rich $\beta$ phase and a washboard-like $\alpha$ phase
are formed. Their formation can be clearly traced back to thermodynamic origins
and can therefore be regarded as inevitable for Pb-substituted Bi cuprates,
independent of the growth method. The $\alpha$- and
$\beta$-phase related features do not show up as expected in LEED for high Pb
content. It is
shown that based on the STM results this LEED picture is not sufficient for
claiming structural purity or complete superstructure suppression.
\begin{acknowledgments} 
We gratefully acknowledge the assistance of H.\,Dwelk and S.\,Rogaschewski in
the characterization of the crystals. We thank V.P.\,Martovitsky for stimulating
discussions. We thank Z.\,Galazka and C.\,\'O Coile\'ain for critical reading
of the manuscript. For
processing the STM data, we used the WSXM software package (Ref.\,\cite{Horcas2007}). 
\end{acknowledgments} 
\bibliography{Luebben2010v2}
\end{document}